\documentclass{IEEEtran}
\usepackage{lineno}
\usepackage{amssymb}
\usepackage{amsfonts}
\usepackage{amsmath}
\usepackage{algorithm}
\usepackage{algorithmic}
\usepackage{tikz}
\usepackage{graphicx,subfigure,cite,multicol,multirow,diagbox,booktabs,array}
\usepackage{color}
\usepackage{bm}
\usepackage{stfloats}

\usepackage{mathtools, cuted}
\usepackage[english]{babel}
\usepackage{lipsum}

\usetikzlibrary{shapes.geometric, arrows}
\ifCLASSINFOpdf
\else
\fi

\begin{document}
\bibliographystyle{IEEEtran}
\title{Precoding and Beamforming Design for Intelligent Reconfigurable Surface-Aided Hybrid Secure Spatial Modulation}

\author{Feng Shu,~Lili Yang,~Yan Wang,~Xuehui Wang,~Weiping~Shi,\\~Chong Shen,~Jiangzhou Wang,~\emph{Fellow},~\emph{IEEE}
\thanks{This work was supported in part by the National Natural Science Foundation of China (Nos. 62071234, 62071289, and 61972093), the Hainan Major Projects (ZDKJ2021022), the Scientific Research Fund Project of Hainan University under Grant KYQD(ZR)-21008, and the National Key R\&D Program of China under Grant 2018YFB180110 (Corresponding authors: Feng Shu).}
\thanks{Feng Shu, Yan Wang, Xuehui Wang and Chong Shen are with the School of Information and Communication Engineering, Hainan University, Haikou, 570228, China(E-mail: shufeng0101@163.com).}
\thanks{Feng Shu, Lili Yang, Weiping Shi are with School of Electronic and Optical Engineering, Nanjing University of Science and Technology, Nanjing, 210094, China.}
\thanks{Jiangzhou Wang is with the School of Engineering and Digital Arts, University of Kent, Canterbury CT2 7NT, U.K. (E-mail: j.z.wang@kent.ac.uk)}}

\maketitle

\begin{abstract}
Intelligent reflecting surface (IRS) is an emerging technology for wireless communication composed of a large number of low-cost passive devices with reconfigurable parameters, which can reflect signals with a certain phase shift and is capable of building programmable communication environment. In this paper, to avoid the high hardware cost and energy consumption in spatial modulation (SM), an IRS-aided hybrid secure SM (SSM) system with a hybrid precoder is proposed. To improve the security performance, we formulate an optimization problem to maximize the secrecy rate (SR) by jointly optimizing the beamforming at IRS and hybrid precoding at the transmitter. Considering that the SR has no closed form expression, an approximate SR (ASR) expression is derived as the objective function. To improve the SR performance, three IRS beamforming methods, called IRS alternating direction method of multipliers (IRS-ADMM), IRS block coordinate ascend (IRS-BCA) and IRS semi-definite relaxation (IRS-SDR), are proposed. As for the hybrid precoding design, approximated secrecy rate-successive convex approximation (ASR-SCA) method and cut-off rate-gradient ascend (COR-GA) method are proposed. Simulation results demonstrate that the proposed IRS-SDR and IRS-ADMM beamformers harvest substantial SR performance gains over IRS-BCA. Particularly, the proposed IRS-ADMM and IRS-BCA are of low-complexity at the expense of a little performance loss compared with IRS-SDR. For hybrid precoding, the proposed ASR-SCA performs better than COR-GA in the high transmit power region.
\end{abstract}


\section{Introduction}{
With the development of modern mobile communication technology, the demand for high-speed data transmission in wireless communication networks is also increasing. Multiple-input multiple-output (MIMO) technology has been a hot topic of research that can significantly improve the performance of wireless communication system \cite{8714426}. Emerging as a new type of MIMO, the basic concept of spatial modulation (SM) technology is to transmit information by utilizing the transmit antenna index and amplitude phase modulation (APM) symbols \cite{4382913}. As SM uses antenna indexes for transmission and only one radio frequency (RF) chain is used,
the design complexity of the system and the detection complexity of the receiving end are greatly simplified \cite{4601434},\cite{4149911}, and the problems of inter-channel interference (ICI) and inter-antenna synchronization (IAS) are effectively avoided \cite{6094024}. Thus, it can achieve a good balance between energy efficiency and spectral efficiency in wireless information transmission.

Due to the broadcast characteristics of wireless channels, the transmitted confidential message (CM) will be intercepted or eavesdropped on by illegal receivers. Traditional upper-layer encryption technology requires strict key distribution technology. However, with the enhancement of computer computing power, this kind of encryption method is also easy to be cracked by a receiver with strong computing power.
In order to ensure the safe transmission of the SM system, physical layer security has been widely considered \cite{6784532,6187751}. There are several ways to improve the security performance of SM, including precoding design \cite{7128369,5956573}, transmit antenna selection \cite{6423761,9829376}, power allocation \cite{9064699,8668810}, and so on.
In \cite{7445202}, a precoding-aided SM (PSM) scheme was proposed to improve the security performance of the system and interfere with the eavesdropper's signal detection.
In \cite{7116516}, authors projected the artificial noise (AN) onto the null-space of the desired channel to enhance security, and two power allocation strategies between CM and AN were proposed in \cite{8730474}.
In addition, the authors in \cite{8373751} confirmed that antenna selection at the transmitter can significantly improve the security performance of the SM system. In  \cite{6663755}, a low-complexity transmit antenna selection method based on matrix singular value decomposition (SVD) was proposed, which had a lower computational complexity to ensure performance improvement compared with an exhaustive search.

The above studies considered full digital (FD) SM.
With the significant increase in the number of transmit antennas, FD SM will cause high hardware costs and energy consumption.
To deal with this dilemma, hybrid SM can be taken into consideration.
The SM transmitter adopting the hybrid structure is composed of two parts, a digital precoder and an analog precoder.
The signal is phase and amplitude modulated by the digital precoder, and then transmitted to the analog precoder by a small amount of RF links. The spatial bit part of the signal in the hybrid SM system will be mapped to the index of the antenna subarray rather than to the single transmit antenna.

In general, transmitters using hybrid structures can be divided into the fully-connected structure and partially-connected structure.
For the fully-connected structure, each RF link is combined with all transmit antennas. This means the phase-modulated signal is transmitted to all transmit antennas, providing sufficient beamforming gain for each RF link \cite{6847111}.
As for the partial-connected structure, each RF link is connected to an antenna subarray, that is, the signal is transmitted through the RF link to the antenna subarray after phase modulation, which means that part of the beamforming gain is sacrificed to a certain extent, thereby greatly reducing the complexity of hardware implementation. Based on its low-complexity hardware property, the partially-connected architecture provides higher energy efficiency than the fully-connected architecture which has a relatively large number of RF chains on the transceiver \cite{7397861}.
The authors in \cite{9296320} combined the hybrid structure transmission with the generalized SM (GSM) system, and proposed a hybrid precoding optimization algorithm, which obtained superior data transmission performance compared with traditional precoding algorithms. Some scholars proposed to combine the hybrid structure with the SM system to improve the symbol error rate (SER) performance of the system \cite{7792630}. In addition, \cite{9262009} studied the antenna selection scheme and the hybrid precoding optimization strategy in the hybrid SM system, and the simulation results showed that the optimization strategy significantly improved the security rate (SR) value of the system.

Intelligent reflection surface (IRS) has gradually become a key technology in the field of sixth generation (6G) research due to its low cost, low power consumption, and programmable characteristics \cite{9475160}. As a revolutionary communication technology, IRS composes of a large number of low-cost passive components with reconfigurable parameters, and it can reflect signals with a certain phase shift without complex precoding \cite{9385923}.
By properly adjusting the phase shift of all passive units at the IRS, the reflected signal can be coherently added in the desired receive direction, or suppressed in the eavesdropping direction to interfere with malicious eavesdroppers, thereby enhancing the security and privacy of the system \cite{9288742}.
Several innovative studies have been carried out on IRS-aided wireless communication systems by jointly optimizing transmit beamforming vector and phase shifts at the IRS \cite{8723525,8743496,9090356,9384498}. An IRS-aided multiple input single output (MISO) model was studied in \cite{8723525}, and the authors designed a solution based on alternating iterative optimization and semidefinite relaxation to obtain a high-quality suboptimal solution.
Additionally, the authors in \cite{8743496} jointly optimized the transmission covariance matrix and the phase shift matrix of IRS to maximize the SR of the MIMO system, and extended the single-antenna eavesdropper to a multi-antenna eavesdropper.

As a special MIMO technology, SM uses an RF chain to transmit information, activates only one transmit antenna at each symbol time, and uses the transmit antenna index to carry additional spatial bit information. Therefore, the combination of IRS and SM can better improve the performance of wireless communication systems and is expected to break the uncontrollability of traditional wireless channels and build a programmable wireless communication environment.
In \cite{9481970}, the authors applied IRS into the SM system and considered two application models, the IRS-aided transmit SM (TSM) and the IRS-aided receive SM (RSM) system. For these two models,  IRS phase shifts optimization algorithms and power allocation algorithms were designed. The results showed that with the assistance of the IRS, compared with the traditional SM system, the SER performance of the system can be well improved. Some scholars also suggested deploying SM technology on the IRS, dividing the reflection units into several areas, and only activating one or more areas at a time to complete the signal reflection, so as to achieve the purpose of transmitting additional information \cite{9769918,9535453,9217944}. The authors in \cite{9133588} extended the IRS-aided RSM to combine both the transmit and receive antenna indexes for joint SM by optimizing the beamforming in IRS.

In the above IRS-aided SM investigations, the high hardware cost and energy consumption caused by many transmit antennas were not considered. To deal with this problem, we introduce the partially-connected hybrid transmit structure into the IRS-aided SM model and propose an IRS-aided hybrid secure SM (SSM) system.
In addition, the signal received by the receiver includes not only the direct signal from the transmitter to the receiver but also the signal forwarded by the IRS. This setting reduces the system hardware as well as improves the transmission performance of the system.
Due to the existence of illegal eavesdroppers, we optimize the beamforming at IRS and transmit hybrid precoding to improve safety performance. The main contribution of this paper are summarized as follows:
\begin{enumerate}
\item  In order to enhance the safety performance of traditional SM system, an IRS-aided hybrid secure SM system model is established, where the partially-connected structure is adopted at the transmitter. Since the transmit antennas have been divided into antenna subarrays, the spatial bits are carried by activating one of the antenna subarrays rather than a single transmit/receive antenna, while the APM symbols are transmitted through the transmit antenna subarray. In addition, the signal received at the receiver consists of the direct signal from the transmitter and the reflected signal from the IRS. Each antenna subarray transmits bit stream by multiple antennas with hybrid precoding and IRS reflects the signal by multiple phase shift units with secure beamforming. These will be utilized to improve the security performance of the IRS-aided hybrid SSM.

\item  In order to improve the security performance of the system, we first derive a cut-off rate based approximate safe rate (ASR) expression as the objective function, and formulate an optimization problem with the aim of maximizing the SR subject to constraints of precoding power limit and unit modulus of IRS reflection elements. For the IRS beamforming design, three beamforming design methods, called IRS alternating direction method of multipliers (IRS-ADMM), IRS block coordinate ascend (IRS-BCA) and IRS semi-definite relaxation (IRS-SDR), are proposed. Simulation results show that the SR performance of the proposed IRS-SDR is better than the proposed IRS-ADMM and IRS-BCA. In particular, compared with the IRS-BCA with a closed-form solution, IRS-ADMM and IRS-SDR harvest higher SR performance but with higher computational complexity.

\item  In order to further improve the security performance of the system, two hybrid precoding methods are proposed based on the ASR expression and cut-off rate (COR) expression, respectively. The approximated secrecy rate-successive convex approximation (ASR-SCA) method is proposed based on the ASR expression. As the expression is difficult to solve directly, we transform the original problem into a concave problem through SCA first and then solve it through optimization. To compare security performance and provide a new solution to this non-convex optimization problem, the cut-off rate-gradient ascend (COR-GA) method is proposed. Simulation results show that both the two proposed hybrid precoding design methods can significantly improve the SR performance, and  the security performance of ASR-SCA is better than that of COR-GA in the high transmit power region.
\end{enumerate}
\emph{Notations:} Boldface lower case and upper case letters denote vectors and matrices, respectively. $(\cdot)^{H}$ denotes the conjugate transpose operation.  $\mathbb{E}\{\cdot\}$ represents expectation operation. $\|\cdot\|$ denotes 2-norm. $\hat{[\;]}$ represents the estimation operation. $\textbf{A}'$ represents a matrix that is different from the original matrix $\textbf{A}$ but has a linear transformation relationship with the original matrix.
}

\section{System Model}
\begin{figure*}[htbp]
\centering
\includegraphics[width=0.9\linewidth]{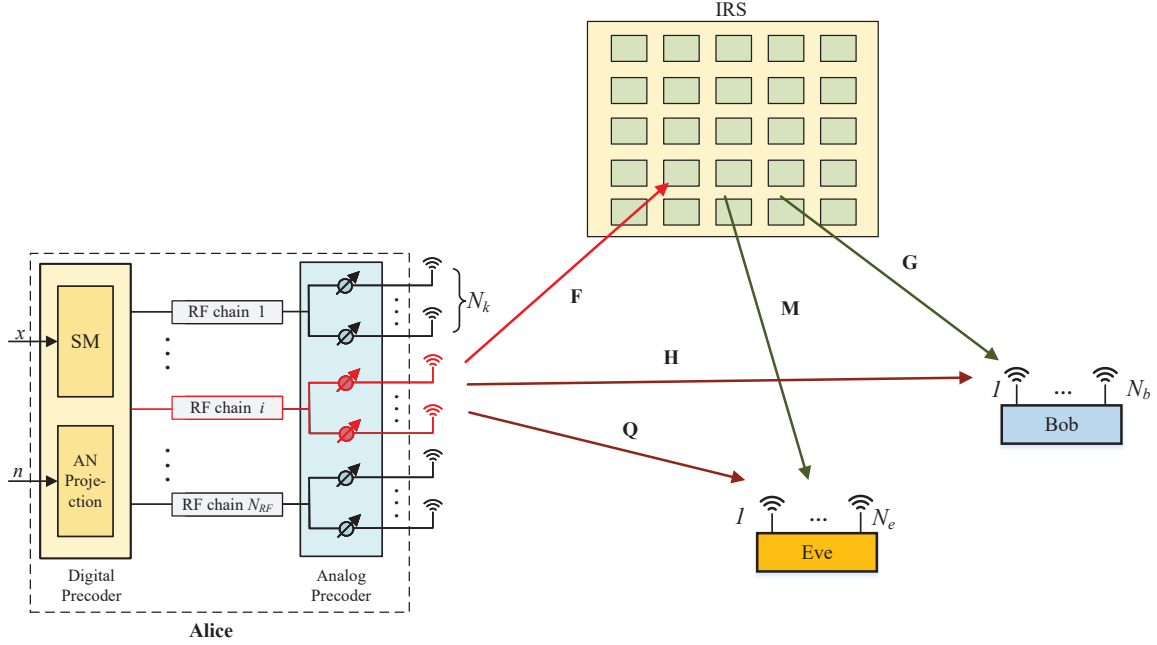}
\centering
\caption{System model for the IRS-aided hybrid secure spatial modulation.}\label{Spatial modulation_fig1}
\end{figure*}
Fig.~\ref{Spatial modulation_fig1} depicts an IRS-aided hybrid SM system consisting of a transmitter (Alice) equipped with partially-connected architecture having $N_{RF}$ RF chains with each owning $N_{k}$ transmit antennas, a desired receiver (Bob) with $N_{b}$ antennas, an illegal eavesdropper receiver (Eve) with $N_{e}$ antennas and an IRS with $N$ low-cost passive reflecting elements.

In this paper, the conventional SM is extended in the hybrid MIMO system. The transmit message is first going through the SM module, and the information is divided into two parts: a symbol chosen from \emph{M}-ary constellation diagram and an index of an RF chain.
In this system, the spatial bits are mapped to the antenna subarrays rather than the single transmit antenna. In order to improve the security performance of the system, AN is also introduced in the transmitter to interfere with the reception of eavesdropping users.
The signal transmitted from Alice is given by
\begin{equation}\label{signal1}
{\bf{x}}_a=\sqrt{\beta P_t}{\mathbf E_i}{\mathbf{F}}_{A}{\mathbf{F}}_{D}{{b}}_{j}+\sqrt{(1-\beta) P_t}{\bf{F}}_{A}{\bf{T}}_{AN}\mathbf{n}_{A}
\end{equation}
where $\beta$ denotes the power allocation factor between CM and AN, $\emph P_t$ denotes the transmit power of CM , and $\mathbf E_i = \text{diag}[\mathbf0,\cdots,\mathbf0,\mathbf{I}_{k},\cdots,\mathbf0] \in \mathbb{R}^{N_{RF}N_k\times N_{RF}N_k}$ is the transmit subarray selection matrix with the $i$th diagonal submatrix being identity matrix and means that the $i$th subarray is chosen to transmit symbol, where $i\in \{1,2,\dots ,N_{RF}\}$. $b_j$ is the $j$th input symbol from the \emph{M}-ary signal constellation,  where $j\in \{1,2\dots M\}$.
$\mathbf F_D \in \mathbb{R}^{N_{RF}\times 1}$ is the digital precoding matrix of CM and has the following expression

\begin{equation}\label{fd}
  \mathbf F_D = [p_1,p_2,\cdots,p_{RF}]^{H}
\end{equation}
After passing through the selected RF chain, the signal will be transmitted to the analog precoder and processed by PS, which is used to phase modulate the signal. The analog precoding matrix $\mathbf{F}_{A}$ is given by
\begin{align}\label{fa}
\mathbf{F}_{A}=\left[\begin{array}{cccc}
\mathbf{f}_{1} & \mathbf{0} & \cdots & \mathbf{0} \\
\mathbf{0} & \mathbf{f}_{2} & \cdots & \mathbf{0} \\
\vdots & \vdots & \ddots & \vdots \\
\mathbf{0} & \mathbf{0} & \cdots & \mathbf{f}_{N_{\mathrm{RF}}}
\end{array}\right]_{N_{RF}N_k \times N_{\mathrm{RF}}}
\end{align}
where $\mathbf{f}_{i}\in \mathbb{C}^{N_{k}\times 1}$ is the analog precoding vector for the $i$th antenna subarray, and all of its elements have the same amplitude $\frac{1}{\sqrt{N_{k}}}$.
$\mathbf T_{AN} \in \mathbb{C}^{N_{RF}\times N_{RF}}$ represents the AN projection matrix, $\mathbf n_{A}\sim {\mathbb{CN} (0,\sigma_A^2\mathbf I_{Nt})}$ is the random AN vector.

The received signal at Bob and Eve can be represented as
\begin{align}\label{signal2}
{\bf{y}}_b=&\sqrt{\beta P_s}({\mathbf {H+GVF}}){\mathbf E_i}{\mathbf{F}}_{A}{\mathbf{F}}_{D}{\text{b}}_{j}+\\
\nonumber        &\sqrt{(1-\beta) P_s}({\mathbf {H+GVF}}){\bf{F}}_{A}{\bf{T}}_{AN}{\bf{n}}+{\mathbf n_B},
\end{align}

\begin{align}\label{signal3}
{\bf{y}}_e=&\sqrt{\beta P_s}({\mathbf {Q+MVF}}){\mathbf E_i}{\mathbf{F}}_{A}{\mathbf{F}}_{D}{\text{b}}_{j}+\\
\nonumber        &\sqrt{(1-\beta) P_s}({\mathbf {Q+MVF}}){\bf{F}}_{A}{\bf{T}}_{AN}{\bf{n}}+{\mathbf n}_E,
\end{align}
where $\mathbf H \in \mathbb{C}^{N_b\times N_{RF}N_k}$,$\mathbf Q \in \mathbb{C}^{N_e\times N_{RF}N_k}$,$\mathbf F \in \mathbb{C}^{N\times N_{RF}N_k}$,$\mathbf G \in \mathbb{C}^{N_b\times N}$,$\mathbf M \in \mathbb{C}^{N_e\times N}$ are the channel gain matrices from Alice to Bob,  Alice to Eve,  Alice to IRS,  IRS to Eve, and IRS to Bob respectively. Finally, $\mathbf n_B\sim {\mathbb{CN} (0,\sigma ^2_B\mathbf I_{N_b})}$ and $\mathbf n_E\sim {\mathbb{CN} (0,\sigma ^2_E\mathbf I_{N_e})}$ represent the complex additive white Gaussian noise (AWGN) vectors at Bob and Eve respectively.
$\mathbf V$ is the reflection coefficient matrix of IRS and is expressed as
\begin{equation}\label{}
  \mathbf V=\text{diag}[v_{1}, v_{2},\cdots, v_{N}]=\text{diag}[\mathbf{v}],
\end{equation}
where $v_i=e^{j\theta_i}$ is the $i$th reflection element and satisfies
\begin{equation}\label{}
  \|v_i\|=1,\theta_i\in[0,2\pi).
\end{equation}

Bob adopts maximum likelihood (ML) detection\cite{4601434}, and the detection process is expressed as
\begin{equation}\label{ml}
[\hat{i},\hat{j}]=\underset{i \in N_{RF},j \in \emph{M}}{\arg \min }||{\bf{y}}_b-\sqrt{\beta P_s}({\mathbf {H+GVF}}){\mathbf E_i}{\mathbf{F}}_{A}{\mathbf{F}}_{D}{\text{b}}_{j}||^{2}
\end{equation}

The average SR is given by
\begin{equation}\label{}
  \bar{R}_s=\mathbf{E}_{\mathbf{H,Q,G,M,F}}(I(\mathbf{x}_a;\mathbf{y}_b)-I(\mathbf{x}_a;\mathbf{y}_e))
\end{equation}
Considering that in the received signal, AN and receiver noise are independent of each other, the zero-mean covariance matrices of the interference plus noise at Bob and Eve can be obtained as

\begin{equation}\label{noise1}
\mathbf\Omega_{B}=(1-\beta)P_s\mathbf{C}_{B}+\sigma_B^2\mathbf{I}_{N_b}
\end{equation}
\begin{equation}\label{noise2}
\mathbf\Omega_{E}=(1-\beta)P_s\mathbf{C}_{E}+\sigma_E^2\mathbf{I}_{N_e}
\end{equation}
where
\begin{equation}\label{noise11}
  \mathbf{C}_{B}=\mathbf{(H+GVF)}{\mathbf{F}}_{A}{\mathbf{T}}_{AN}{{\mathbf{T}}_{AN}^{H}}{\mathbf{F}}_{A}^{H}{\mathbf{(H+GVF)}^{H}}
\end{equation}
\begin{equation}\label{noise12}
  \mathbf{C}_{E}=\mathbf{(Q+MVF)}{\mathbf{F}}_{A}{\mathbf{T}}_{AN}{{\mathbf{T}}_{AN}^{H}}{\mathbf{F}}_{A}^{H}{\mathbf{(Q+MVF)}^{H}}
\end{equation}
Therefore, without affecting the mutual information, i.e. $I(\mathbf{x}_a;\mathbf{y}_B)=I(\mathbf{x}_a;\mathbf{y}_B^{'})$ and $I(\mathbf{x}_a;\mathbf{y}_E)=I(\mathbf{x}_a;\mathbf{y}_E^{'})$, the AN interference term and receiver noise term are whitened and the received signals can be given by ${\bf{y}}_b^{'}=\mathbf\Omega_{b}^{-\frac{1}{2}}{\bf{y}}_b$ and ${\bf{y}}_e^{'}=\mathbf\Omega_{e}^{-\frac{1}{2}}{\bf{y}}_e$. Therefore, the average SR is rewritten as $\bar{R}_s=\mathbf{E}_{\mathbf{H,Q,G,M,F}}(I(\mathbf{x}_a;\mathbf{y}_b^{'})-I(\mathbf{x}_a;\mathbf{y}_e^{'}))$.

For the convenience of expression, a new variable $X_k$, where $k\in\{1,2,\cdots,MN_{RF}\}$ is introduced, which denotes one of the $MN_{RF}$ possibilities of the transmit signal $\mathbf x_a$,
\begin{equation}\label{CM}
  \mathbf{X}_k={\mathbf E_i}{b}_{j}
\end{equation}

Thus, the mutual information between the transmit signal and $\mathbf {y}_b^{'}$ can be expressed as:
\begin{align}\label{signal5}
I(\mathbf{x}_a;{\bf{y}}_b^{'})&={\rm{log}}_2N_{RF}M-\frac{1}{N_{RF}M}{\sum_{m=1}^{N_{RF}M}}\\
\nonumber    &{\mathbb{E}_{\mathbf{n_{B}^{'}}}}({\rm{log}}_2\sum_{n=1}^{N_{RF}M}
       {\rm{exp}}[-\frac{||f_{B,m,n}+\mathbf n_{B}^{'}||^{2}+||\mathbf n_{B}^{'}||^{2}}{\sigma^{2}}]),
\end{align}
where
\begin{equation}\label{}
  f_{B,m,n}=\sqrt{\beta P_s}\mathbf\Omega_{B}^{-\frac{1}{2}}\mathbf{(H+GVF)}(\mathbf X_{m}-\mathbf X_{n})\mathbf{F}_{A}{\mathbf{F}}_{D}.
\end{equation}

Similarly, the mutual information between the transmit signal and $\mathbf {y}_e^{'}$ can be expressed as:
\begin{align}\label{signal6}
I(\mathbf{x}_a;{\bf{y}}_e^{'})&={\rm{log}}_2N_{RF}M-\frac{1}{N_{RF}M}{\sum_{i=1}^{N_{RF}M}}\\
\nonumber         &{\mathbb{E}_{\mathbf{n_{E}^{'}}}}({\rm{log}}_2\sum_{j=1}^{N_{RF}M}
       {\rm{exp}}[-\frac{||f_{E,i,j}+\mathbf n_{E}^{'}||^{2}+||\mathbf n_{E}^{'}||^{2}}{\sigma^{2}}])
\end{align}
where
\begin{equation}\label{}
  f_{E,i,j}=\sqrt{\beta P_s}\mathbf\Omega_{E}^{-\frac{1}{2}}\mathbf{(Q+MVF)}(\mathbf X_{i}-\mathbf X_{j})\mathbf{F}_{A}{\mathbf{F}}_{D}
\end{equation}

In order to improve the safety performance of the system, the following research is mainly based on maximizing the SR. Using the above expression causes extremely high computational complexity and requires a large number of Monte Carlo simulations, which is not conducive to the optimization design. The cut-off rate has proven to be very close to the Monte Carlo-based SR in \cite{8638547}.
In accordance with the definition of the cut-off rate for traditional MIMO systems in \cite{7891030}, the cut-off rates for Bob and Eve are expressed as
\begin{align}\label{cutt1}
I^{B}_{0}=&2{\rm{log}_{2}}N_{RF}M-{\rm{log}_2}\sum_{m=1}^{N_{RF}M}\sum_{n=1}^{N_{RF}M}\\
\nonumber  &{\rm{exp}}(\frac{-||\sqrt{\beta P_s}\mathbf\Omega_{B}^{-\frac{1}{2}}{\bf{(H+GVF)}}\mathbf D_{mn}\mathbf{F}_{A}{\mathbf{F}}_{D}||^2}{4})
\end{align}
and
\begin{align}\label{cutt2}
  I^{E}_{0}=&2{\rm{log}_{2}}N_{RF}M-{\rm{log}_2}\sum_{m^{'}=1}^{N_{RF}M}\sum_{n^{'}=1}^{N_{RF}M}\\
\nonumber  &{\rm{exp}}(\frac{-||\sqrt{\beta P_s}\mathbf\Omega_{E}^{-\frac{1}{2}}{\bf{(Q+MVF)}}\mathbf D_{m^{'}n^{'}}\mathbf{F}_{A}{\mathbf{F}}_{D}||^2}{4}),
\end{align}
, respectively, where
\begin{equation}\label{}
  \mathbf D_{mn}=\mathbf X_m-\mathbf X_n
\end{equation}
and
\begin{equation}\label{}
  \mathbf D_{m^{'}n^{'}}=\mathbf X_m^{'}-\mathbf X_n^{'}.
\end{equation}

Therefore, the average SR for the system can be formulated as
\begin{align}\label{rs}
R^{a}_{s}={\rm{log}_2}\kappa_{E}-{\rm{log}_2}\kappa_{B}
\end{align}
where
\begin{align}\label{}
  \kappa_{E}=&\sum_{m^{'}=1}^{N_{RF}M}\sum_{n^{'}=1}^{N_{RF}M}\\
\nonumber &{\rm{exp}}(\frac{-||\sqrt{\beta P_s}\mathbf\Omega_{E}^{-\frac{1}{2}}{\bf{(Q+MVF)}}\mathbf D_{m^{'}n^{'}}\mathbf{F}_{A}{\mathbf{F}}_{D}||^2}{4}),
\end{align}
\begin{align}\label{}
 \kappa_{B}=&\sum_{m=1}^{N_{RF}M}\sum_{n=1}^{N_{RF}M}\\
\nonumber &{\rm{exp}}(\frac{-||\sqrt{\beta P_s}\mathbf\Omega_{B}^{-\frac{1}{2}}{\bf{(H+GVF)}}\mathbf D_{mn}\mathbf{F}_{A}{\mathbf{F}}_{D}||^2}{4}).
\end{align}
 Although the SR expression has been approximated by cut-off rate, there are still three variables to be optimized in Eq.(\ref{rs}), which are $\bf V$, $\mathbf F_{A}$ and $\mathbf F_{D}$ respectively. In order to simplify the optimization problem, we define the hybrid precoding matrix $\mathbf{P}$ as $\mathbf{P}=\mathbf{F}_{A}{\mathbf{F}}_{D}$, i.e.
 \begin{equation}\label{}
   \mathbf{P}=[\mathbf{p}_1^{T},\mathbf{p}_2^{T},\cdots,\mathbf{p}_{RF}^{T}]^{T},
 \end{equation}
 where $\mathbf{p}_i\in\mathbb C^{N_k\times1}$,$i=1,2,\cdots,N_{RF}$.

Finally, our objective is to maximize the SR by designing the beamforming at IRS and the transmit hybrid precoding at Alice, which is casted as the following optimization problem
\begin{align}
\left( \textbf{P1} \right):	\underset{\mathbf P~\mathbf V}{\mathrm {max}}  \quad &{R^{a}_{s}}(\mathbf P, \mathbf V) \label{main}  \\
	\mathrm{s.t.}  \quad  &||\mathbf P||\leq N_{RF} \tag{\ref{main}{a}} \label{maina}\\
                           &||\mathbf V_{n,n}||=1 \tag{\ref{main}{b}} \label{mainb}
\end{align}
The optimization problem (\ref{main}) is difficult to solve as $\mathbf P$ and $\mathbf V$ are coupled and the objective function is non-concave, but we can split the objective function into two sub-problems by fixing one of $\mathbf P$ and $\mathbf V$. Thus, the objective function \textbf{P1} can be solve by alternatively optimizing the subproblems \textbf{P1-1} and \textbf{P1-2}.
\section{Proposed IRS beamforming methods}
{For the IRS-aided hybrid SM system, the design of IRS beamformer is necessary to improve the system performance. In this section, three IRS beamformers IRS strategies, called IRS-ADMM, IRS-BCD and IRS-SDR, are proposed to enhance the security performance of IRS-aided hybrid SM system.

We rewrite \textbf{P1} into  \textbf{P1-1} with $\mathbf{P}$ fixed,
\begin{align}
\left( \textbf{P1-1} \right):	\underset{\mathbf V}{\mathrm {max}}  \quad &{R^{a}_{s}}(\mathbf V) \label{fxp}  \\
	\mathrm{s.t.}  \quad   &||\mathbf V_{n,n}||=1 \tag{\ref{fxp}{a}} \label{fxpa}
\end{align}
We can transform the numerators in \textbf{P1-1} into the following forms:
\begin{align}\label{1trans}
&||\sqrt{\beta P_s}\mathbf\Omega_{B}^{-\frac{1}{2}}{\bf{(H+GVF)}}\mathbf D_{mn}\mathbf{P}||^2=\\
\nonumber&||\sqrt{\beta P_s}{\bf{(\widetilde{H}+\widetilde{G}VF)}}\mathbf D_{mn}\mathbf{P}||^2,
\end{align}
\begin{align}\label{11trans}
&||\sqrt{\beta P_s}\mathbf\Omega_{E}^{-\frac{1}{2}}{\bf{(Q+MVF)}}\mathbf D_{m^{'}n^{'}}\mathbf{P}||^2=\\
\nonumber&||\sqrt{\beta P_s}{\bf{(\widetilde{Q}+\widetilde{M}VF)}}\mathbf D_{m^{'}n^{'}}\mathbf{P}||^2,
\end{align}
where
\begin{align}\label{}
  &\mathbf\Omega_{B}^{-\frac{1}{2}}\mathbf{H}=\widetilde{\bf H}\\
\nonumber  &\mathbf\Omega_{B}^{-\frac{1}{2}}\mathbf{G}=\widetilde{\bf G},
\end{align}
and
\begin{align}\label{}
  &\mathbf\Omega_{E}^{-\frac{1}{2}}\mathbf{Q}=\widetilde{\bf Q}\\
\nonumber  &\mathbf\Omega_{E}^{-\frac{1}{2}}\mathbf{M}=\widetilde{\bf M}.
\end{align}

As the equations $\mathbf{D}_{mn}=\mathbf{E}_{i}b_{j}-\mathbf{E}_{m}b_{n}$ and $\mathbf{D}_{m^{'}n^{'}}=\mathbf{E}_{i^{'}}b_{j^{'}}-\mathbf{E}_{m^{'}}b_{n^{'}}$ hold, where $\mathbf{E}_{i}$ means choosing the $i$th subarray to convey information, the precoding matrix can be as the production of Eq.(\ref{fd}) and Eq.(\ref{fa}), therefore,  Eq.(\ref{1trans}) can be rewritten as:
\begin{equation}\label{2trans}
  \beta P_s||(\widetilde{\mathbf{H}}_{i}+\widetilde{\mathbf{G}}\mathbf{V}{\mathbf{F}_{i}}){p_{i}}{\mathbf{f}_{i}}{b_{j}}-(\widetilde{\mathbf{H}}_{m}+\widetilde{\mathbf{G}}\mathbf{V}{\mathbf{F}_{m}}){p_{m}}{\mathbf{f}_{m}}{b_{n}}||^{2}.
\end{equation}
Observing this equation, we denote $p_{i}\mathbf{F}_{i}\mathbf{f}_{i}b_{j}\in \mathbb{C}^{N\times 1}$ and $p_{i}\widetilde{\mathbf{H}}_{i}\mathbf{f}_{i}b_{j}\in \mathbb{C}^{N\times 1}$ as vectors $\mathbf{s}_{ij}$ and $\mathbf{a}_{ij}$, and Eq.(\ref{2trans}) can be expressed as:
\begin{align}\label{asij}
   &\beta P_s||(\mathbf{a}_{ij}+\widetilde{\mathbf{G}}\mathbf{V}\mathbf{s}_{ij})-(\mathbf{a}_{mn}+\widetilde{\mathbf{G}}\mathbf{V}\mathbf{s}_{mn})||^{2}\\
 \nonumber  &=\beta P_s||(\mathbf{a}_{ij}+\widetilde{\mathbf{G}}\mathbf{S}_{ij}\mathbf{v})-(\mathbf{a}_{mn}+\widetilde{\mathbf{G}}\mathbf{S}_{mn}\mathbf{v})||^{2}\\
 \nonumber  &=\beta P_s||(\mathbf{a}_{ij}-\mathbf{a}_{mn})+\widetilde{\mathbf{G}}(\mathbf{S}_{ij}-\mathbf{S}_{mn})\mathbf{v}||^{2},
\end{align}
where $\mathbf{S}_{ij}=\text{diag}(\mathbf{s}_{ij})$, and then we can turn (\ref{1trans}) into a quadratic form:
\begin{align}\label{quadratic1}
  &||(\mathbf{a}_{ij}-\mathbf{a}_{mn})+\widetilde{\mathbf{G}}(\mathbf{S}_{ij}-\mathbf{S}_{mn})\mathbf{v}||^{2}\\
\nonumber   &=\mathbf{v}^{H}\mathbf{B}^{mn}_{ij}\mathbf{v}+2\mathfrak{Re}\{(\mathbf{A}^{mn}_{ij})^{H}\mathbf{C}^{i,j}_{m,n}\mathbf{v}\}+||\mathbf{A}^{mn}_{ij}||^{2},
\end{align}
where
\begin{equation}\label{}
  \mathbf{A}^{mn}_{ij}=\mathbf{a}_{ij}-\mathbf{a}_{mn},
\end{equation}
\begin{equation}\label{}
  \mathbf{B}^{mn}_{ij}=[\widetilde{\mathbf{G}}(\mathbf{S}_{ij}-\mathbf{S}_{mn})]^{H}[\widetilde{\mathbf{G}}(\mathbf{S}_{ij}-\mathbf{S}_{mn})],
\end{equation}
\begin{equation}\label{}
  \mathbf{C}^{i,j}_{m,n}=\widetilde{\mathbf{G}}(\mathbf{S}_{ij}-\mathbf{S}_{mn}).
\end{equation}
}
In the same way, we can turn the (\ref{11trans}) into the following quadratic form:
\begin{align}\label{}
 &||{\bf{(\widetilde{Q}+\widetilde{M}VF)}}\mathbf D_{m^{'}n^{'}}\mathbf{P}||^{2}\\
\nonumber   &=\mathbf{v}^{H}\mathbf{B}^{m^{'}n^{'}}_{i^{'}j^{'}}\mathbf{v}+2\mathfrak{Re}\{(\mathbf{A}^{m^{'}n^{'}}_{i^{'}j^{'}})^{H}\mathbf{C}^{m^{'}n^{'}}_{i^{'}j^{'}}\mathbf{v}\}+||\mathbf{A}^{m^{'}n^{'}}_{i^{'}j^{'}}||^{2}.
\end{align}

In this way, the objective function in problem (\textbf{P1-1}) is converted into
\begin{align}\label{}
   R^{a}_{s}=&{\rm{log}_2}\sum_{i^{'},m^{'}=1}^{N_{RF}}\sum_{j^{'},n^{'}=1}^{M}{\rm{exp}}(-\tau(\mathbf{v}^{H}\mathbf{B}^{m^{'}n^{'}}_{i^{'}j^{'}}\mathbf{v}+\\
 \nonumber &2\mathfrak{Re}\{(\mathbf{A}^{m^{'}n^{'}}_{i^{'}j^{'}})^{H}\mathbf{C}^{m^{'}n^{'}}_{i^{'}j^{'}}\mathbf{v}\}+||\mathbf{A}^{m^{'}n^{'}}_{i^{'}j^{'}}||^{2}))-\\
 \nonumber  &{\rm{log}_2}\sum_{i,m=1}^{N_{RF}}\sum_{j,n=1}^{M}{\rm{exp}}(-\tau(\mathbf{v}^{H}\mathbf{B}^{mn}_{ij}\mathbf{v}+\\
 \nonumber &2\mathfrak{Re}\{(\mathbf{A}^{mn}_{ij})^{H}\mathbf{C}^{mn}_{ij}\mathbf{v}\}+||\mathbf{A}^{mn}_{ij}||^{2})).
\end{align}
Using Jensen's inequality, we can get the lower bound of $R^{a}_{s}$ as
\begin{align}\label{rl}
R^{l}_{s}=&{\rm{log}_2 e}\left[\sum_{i^{'},m^{'}=1}^{N_{RF}}\sum_{j^{'},n^{'}=1}^{M}\bigg(-\tau\Big(\mathbf{v}^{H}\mathbf{B}^{m^{'}n^{'}}_{i^{'}j^{'}}\mathbf{v}+\right.\\
 \nonumber &\left.2\mathfrak{Re}\{(\mathbf{A}^{m^{'}n^{'}}_{i^{'}j^{'}})^{H}\mathbf{C}^{m^{'}n^{'}}_{i^{'}j^{'}}\mathbf{v}\}+||\mathbf{A}^{m^{'}n^{'}}_{i^{'}j^{'}}||^{2}\Big)\bigg)\right]-\\
 \nonumber  &{\rm{log}_2 e}\left[\sum_{i,m=1}^{N_{RF}}\sum_{j,n=1}^{M}\bigg(-\tau\Big(\mathbf{v}^{H}\mathbf{B}^{mn}_{ij}\mathbf{v}+\right.\\
 \nonumber &\left.2\mathfrak{Re}\{(\mathbf{A}^{mn}_{ij})^{H}\mathbf{C}^{mn}_{ij}\mathbf{v}\}+||\mathbf{A}^{mn}_{ij}||^{2}\Big)\bigg)\right].
\end{align}

\subsection{Proposed IRS-ADMM method}
{Thus, based on Eq.(\ref{rl}), problem (\textbf{P1-1}) can be expressed as:
\begin{align}
\left( \textbf{P1-1-1} \right): \nonumber \underset{\mathbf v}{\mathrm {max}}  \quad &\mathbf{v}^{H}\mathbf{\Phi}_{B}\mathbf{v}-\mathbf{v}^{H}\mathbf{\Phi}_{E}\mathbf{v}+2\mathfrak{Re}\{\mathbf{D}\mathbf{v}\}\\
&-2\mathfrak{Re}\{\mathbf{D}^{'}\mathbf{v}\}+\mathbf{C} \label{DC}\\
\mathrm{s.t.}  \quad   &||\mathbf V_{n,n}||=1 \tag{\ref{DC}{a}} \label{DCa}
\end{align}
where
\begin{equation}\label{phib}
 \mathbf{\Phi}_{B}={\rm{log}_2 e}\big(\sum_{i,m=1}^{N_{RF}}\sum_{j,n=1}^{M}\tau\mathbf{B}^{mn}_{ij}\big),
\end{equation}
\begin{equation}\label{phie}
\mathbf{\Phi}_{E}={\rm{log}_2 e}\big(\sum_{i^{'},m^{'}=1}^{N_{RF}}\sum_{j^{'},n^{'}=1}^{M}\tau\mathbf{B}^{m^{'}n^{'}}_{i^{'}j^{'}}\big),
\end{equation}
\begin{equation}\label{d}
\mathbf{D}={\rm{log}_2 e}\Big(\sum_{i,m=1}^{N_{RF}}\sum_{j,n=1}^{M}\big(\tau(\mathbf{A}^{mn}_{ij})^{H}\mathbf{C}^{mn}_{ij}\big)\Big),
\end{equation}
\begin{equation}\label{dd}
\mathbf{D}^{'}={\rm{log}_2 e}\Big(\sum_{i^{'},m^{'}=1}^{N_{RF}}\sum_{j^{'},n^{'}=1}^{M}\big(\tau(\mathbf{A}^{m^{'}n^{'}}_{i^{'}j^{'}})^{H}\mathbf{C}^{m^{'}n^{'}}_{i^{'}j^{'}}\big)\Big),
\end{equation}
and
\begin{equation}\label{c}
\mathbf{C}=\tau[\sum_{i,m=1}^{N_{RF}}\sum_{j,n=1}^{M}||\mathbf{A}^{mn}_{ij}||^{2}-\sum_{i^{'},m^{'}=1}^{N_{RF}}\sum_{j^{'},n^{'}=1}^{M}||\mathbf{A}^{m^{'}n^{'}}_{i^{'}j^{'}}||^{2}].
\end{equation}

Since the objective function in Eq. (\ref{DC}) is a difference of convex (DC) function, we take the first-order Taylor expansion of
 $\mathbf{v}^{H}\mathbf{\Phi}_{B}\mathbf{v}$ at a given feasible point as follows:
\begin{equation}\label{}
  \mathbf{v}^{H}\mathbf{\Phi}_{B}\mathbf{v}\geq2\mathfrak{Re}\{ \mathbf{v}_{o}^{H}\mathbf{\Phi}_{B}\mathbf{v}\}- \mathbf{v}^{H}_{o}\mathbf{\Phi}_{B}\mathbf{v}_{o}.
\end{equation}
and the Eq. (\ref{DC}) turns into:
\begin{align}\label{}
\nonumber \underset{\mathbf v}{\mathrm {max}}  \quad 2 &\mathfrak{Re}\{ \mathbf{v}_{o}^{H}\mathbf{\Phi}_{B}\mathbf{v}\}- \mathbf{v}^{H}_{o}\mathbf{\Phi}_{B}\mathbf{v}_{o}-\mathbf{v}^{H}\mathbf{\Phi}_{E}\mathbf{v}\\
 &+2\mathfrak{Re}\{\mathbf{D}\mathbf{v}\}-2\mathfrak{Re}\{\mathbf{D}^{'}\mathbf{v}\}+\mathbf{C} \label{simple} \\
\mathrm{s.t.}  \quad   &||\mathbf v_{n}||=1 \tag{\ref{simple}{a}} \label{simplea}.
\end{align}
As (\ref{simplea}) is a non-convex constraint, we introduce an auxiliary slack variable $\mathbf{u}$, which satisfies $\mathbf{u}=\mathbf{v}$, as well as a penalty term for $\mathbf{u}\neq\mathbf{v}$  and the objective function turns into:
\begin{align}\label{}
 \nonumber\underset{\mathbf v~\mathbf u}{\mathrm {max}}  \quad &2 \mathfrak{Re}\{ \mathbf{u}_{o}^{H}\mathbf{\Phi}_{B}\mathbf{u}\}- \mathbf{u}^{H}_{o}\mathbf{\Phi}_{B}\mathbf{u}_{o}-\mathbf{u}^{H}\mathbf{\Phi}_{E}\mathbf{u}\\
 &+2\mathfrak{Re}\{\mathbf{D}\mathbf{u}\}-2\mathfrak{Re}\{\mathbf{D}^{'}\mathbf{u}\}+\mathbf{C} \label{simple1} \\
\mathrm{s.t.}  \quad    &\mathbf{u}=\mathbf{v} \tag{\ref{simple1}{a}} \label{simple1a}\\
            &||\mathbf v_{n}||=1 \tag{\ref{simple1}{b}} \label{simple1b}
\end{align}
Omitting the irrelevant term, the augmented Lagrange function is expressed as:
\begin{align}\label{auglag}
\mathcal{L(\mathbf{u},\mathbf{v},\bm{\lambda})}=&2 \mathfrak{Re}\{ \mathbf{u}_{o}^{H}\mathbf{\Phi}_{B}\mathbf{u}\}- \mathbf{u}^{H}_{o}\mathbf{\Phi}_{B}\mathbf{u}_{o}-\mathbf{u}^{H}\mathbf{\Phi}_{E}\mathbf{u}\\
\nonumber&-\frac{\rho}{2}||\mathbf{u}-\mathbf{v}||^{2}+\mathfrak{Re}\{\bm{\lambda}^{H}(\mathbf{u}-\mathbf{v})\}\\
\nonumber&+2\mathfrak{Re}\{\mathbf{D}\mathbf{u}\}-2\mathfrak{Re}\{\mathbf{D}^{'}\mathbf{u}\},
\end{align}
where $\rho$ is the penalty parameter and $\lambda$ is the dual variable for $\mathfrak{Re}\{\bm{\lambda}^{H}(\mathbf{u}-\mathbf{v})\}$.

And we can get the dual Lagrange function as
\begin{equation}\label{duallag}
  \underset{\bm{\lambda}}{\mathrm {min}}\quad\mathcal{D(\bm{\lambda})}=\underset{\mathbf v~\mathbf u}{\mathrm {max}}  \quad\mathcal{L(\mathbf{u},\mathbf{v},\bm{\lambda})}.
\end{equation}
The closed-form solutions can be represented as:
\begin{equation}\label{}
\bm{u}^{n+1}=\arg \max _{\bm{u}} \mathcal{L}\left(\bm{u}^{n}, \bm{v}^{n}, \bm{\lambda}^{n}\right)
\end{equation}
\begin{equation}\label{}
\bm{v}^{n+1}=\arg \max _{\bm{v}} \mathcal{L}\left(\bm{u}^{n+1}, \bm{v}^{n}, \bm{\lambda}^{n}\right)
\end{equation}
\begin{equation}\label{lamb1}
\bm{\lambda}^{n+1}=\bm\lambda^{n}-\rho(\bm{u}^{n+1}-\bm{v}^{n+1})
\end{equation}

In order to solve $\bm{u}^{n+1}$, we derive the first order derivative of Eq.(\ref{auglag}) with respect to $\mathbf{u}$ while fixing $\mathbf{v}$ and $\bm{\lambda}$:
\begin{align}\label{deri}
  2\mathbf{\Phi}_{B}^{H}\mathbf{u}_{o}-&2\mathbf{\Phi}_{E}\mathbf{u}^{n+1}+\bm{\lambda}^{n}-\rho(\mathbf{u}^{n+1}-\mathbf{v}^{n})\\
  \nonumber&+2(\mathbf{D})^{H}-2(\mathbf{D}^{'})^{H}=0,
\end{align}
and we can get the updated formula of $\mathbf{\bm{u}}$ as follows:
\begin{align}\label{admm1}
  \bm{u}^{n+1}=&( 2\mathbf{\Phi}_{E}+\rho\mathbf{I}_{N})^{-1}(2\mathbf{\Phi}_{B}^{H}\mathbf{u}_{o}+\bm{\lambda}^{n}+\\
  \nonumber&\rho\bm{v}^{n}+2(\mathbf{D})^{H}-2(\mathbf{D}^{'})^{H}).
\end{align}
In Eq.(\ref{deri}), $\mathbf{v}$ is optimized with $\mathbf{u}$ and $\bm{\lambda}$ fixed and we have:
\begin{equation}\label{admm3}
[\mathbf{v}^{n+1}]_{m}=\left\{
\begin{array}{cl}
\frac{[(\mathbf{u}^{n+1}-\rho^{-1}\bm{\lambda}^{n})]_{m}}{|[(\mathbf{u}^{n+1}-\rho^{-1}\bm{\lambda}^{n})]_{m}|} & [(\mathbf{u}^{n+1}-\rho^{-1}\bm{\lambda}^{n})]_{m}\neq 0 \\
{{[\mathbf{v}^{n}]}_{m}} & [(\mathbf{u}^{n+1}-\rho^{-1}\bm{\lambda}^{n})]_{m}= 0 \\
\end{array} \right.
\end{equation}

Finally, the updated expression of $\bm{\lambda}$ can be obtained through Eq.(\ref{lamb1}).

Additionally, a step-by-step summary is provided as follows:
$\textbf{Algorithm 1}$
}
\begin{algorithm}
	\renewcommand{\algorithmicrequire}{\textbf{Input:}}
	\renewcommand{\algorithmicensure}{\textbf{Output:}}
	\caption{Proposed IRS-ADMM beamformer}
	\label{alg:1}
	\begin{algorithmic}[1]
		\REQUIRE the channel matrix $\textbf{H}$, $\textbf{Q}$, $\textbf{M}$, $\textbf{G}$, $\textbf{F}$, $\mathbf{P}$, and the $\mathcal{M}$-ary constellation
		\ENSURE $\bm {v}$
        \STATE Computing $\mathbf \Phi_{B}$, $\mathbf \Phi_{E}$, $\mathbf{D}$ and $\mathbf{D}^{'}$ according to (\ref{phib}) to (\ref{dd})
		\STATE Initialize $\mathbf {v}_0$ to a feasible value.
        \STATE Initialize $k=0$
        \STATE Initialize $\mathbf {u}_0$, $\lambda_{0}$ similar to $\mathbf v_{0}$.
        \STATE Reformulate the problem by the ADMM method to get (\ref{duallag})
        \REPEAT
        \STATE Let $k=k+1$
        \REPEAT
        \STATE Update the auxiliary slack variable $\mathbf u_{k}$ by (\ref{admm1})
        \STATE Update the beamforming vector $\mathbf v_{k}$ by (\ref{admm3})
        \STATE Update the dual variable $\lambda_{k}$ by (\ref{lamb1})
        \UNTIL $\|\mathbf v_{k}-\mathbf v_{k-1}\|_2\leq0.01$
        \UNTIL $\|{R^{a}_{s}}(\mathbf v_{k})-{R^{a}_{s}}(\mathbf v_{k-1})\|_2\leq0.01$
        \STATE \textbf{return} $\mathbf v$
	\end{algorithmic}
\end{algorithm}
\subsection{Proposed IRS-BCA method}
In the previous section, the IRS-ADMM algorithm was presented to optimize the secure IRS beamforming vector based on the augmented Lagrange function. For the comparison of the secrecy performance and to offer a new computational complexity solution to this non-convex optimization problem, we propose another secure IRS beamforming method with less computational complexity performance, namely IRS-BCA, in the following.

It is easy to convert the Eq.(\ref{DC}) into the following form:
\begin{align}
\left( \textbf{P1-1-2} \right): \underset{\mathbf v}{\mathrm {max}}  \quad &\mathbf{v}^{H}\mathbf{\Phi}\mathbf{v}+2\mathfrak{Re}\{\mathbf{\Delta}\mathbf{v}\} \label{simple2} \\
\mathrm{s.t.}  \quad   &||\mathbf v_{n}||=1 \tag{\ref{simple2}{a}} \label{simple2a}
\end{align}
where $\mathbf{\Phi}=\mathbf{\Phi}_{B}-\mathbf{\Phi}_{E}$ and $\mathbf{\Delta}=\mathbf{D}-\mathbf{D}^{'}$,

As $\mathbf{v}^{H}\mathbf{\Phi}\mathbf{v}=\sum_{N}^{N}v_{i}^{H}\phi_{i,j} v_{j}$, the objective function can be written as:
\begin{align}\label{bcd}
  2&\mathfrak{Re}\{v^{H}_{n}(\sum_{j\neq n}^{N} \phi_{n,j} v_{j}+\delta_{n})\}+2\mathfrak{Re}\big[\sum_{i\neq n}^{N}v^{H}_{i}\delta_{i}^{H}\big]\\
\nonumber &+\sum_{i\neq n}^{j\neq n}v^{H}_{i}\phi_{i,j}v_{j}+v_{n}^{H}\phi_{n,n}v_{n}.
\end{align}

To maximize the above objective function, the optimal $v_{n}$ for Eq.(\ref{bcd}) can be derived as follows:
\begin{equation}\label{constant}
  v_{n}^{*}=\frac{\sum_{j\neq n}^{N} \phi_{n,j} v_{j}+\delta_{n}^{H}}{|\sum_{j\neq n}^{N} \phi_{n,j} v_{j}+\delta_{n}^{H}|}
\end{equation}

Overall, all the reflection units in IRS can be optimized from $n=1$ to $n=N$ and repeated till convergence is reached.
\subsection{Proposed IRS-SDR method}
In order to solve the problem that the constraint is non-convex in problem \textbf{P1-1}, different from the above two subsections, Eq.(\ref{fxp}) is expressed in a new form in this subsection. The SDR method is utilized to further relax the non-convex constraints and solve the optimization problem.
The numerators in \textbf{P1-1} have other expressions as follows:
\begin{align}\label{sdr1}
&||{\bf{(\widetilde{H}+\widetilde{G}VF)}}\mathbf D_{mn}\mathbf{P}||^{2}\\
\nonumber   &=\mathbf{v}^{H}\mathbf{B}^{mn}_{ij}\mathbf{v}+(\mathbf{A}^{mn}_{ij})^{H}\mathbf{C}^{i,j}_{m,n}\mathbf{v}+\mathbf{v}^{H}(\mathbf{C}^{i,j}_{m,n})^{H}\mathbf{A}^{mn}_{ij})\\
\nonumber   &+||\mathbf{A}^{mn}_{ij}||^{2},
\end{align}
\begin{align}\label{sdr2}
 &||{\bf{(\widetilde{Q}+\widetilde{M}VF)}}\mathbf D_{m^{'}n^{'}}\mathbf{P}||^{2}\\
\nonumber   &=\mathbf{v}^{H}\mathbf{B}^{m^{'}n^{'}}_{i^{'}j^{'}}\mathbf{v}+(\mathbf{A}^{m^{'}n^{'}}_{i^{'}j^{'}})^{H}\mathbf{C}^{m^{'}n^{'}}_{i^{'}j^{'}}\mathbf{v}+\mathbf{v}^{H}(\mathbf{C}^{m^{'}n^{'}}_{i^{'}j^{'}})^{H}\mathbf{A}^{m^{'}n^{'}}_{i^{'}j^{'}}\\
\nonumber   &+||\mathbf{A}^{m^{'}n^{'}}_{i^{'}j^{'}}||^{2}.
\end{align}
Using the Jensen's inequality, the lower bound of (\textbf P1-1) has a new expression as:
\begin{align}
\left( \textbf{P1-1-3} \right): \nonumber \underset{\mathbf v}{\mathrm {max}}  \quad &\mathbf{v}^{H}\mathbf{\Phi}_{B}\mathbf{v}-\mathbf{v}^{H}\mathbf{\Phi}_{E}\mathbf{v}+\mathbf{D}\mathbf{v}+\mathbf{v}^{H}\mathbf{D}^{H}\\
&-\mathbf{D}^{'}\mathbf{v}-\mathbf{v}^{H}\mathbf{D}^{'H}+\mathbf{C} \label{DC1}\\
\mathrm{s.t.}  \quad   &||\mathbf V_{n,n}||=1. \tag{\ref{DC1}{a}} \label{DC1a}
\end{align}

After simplification and omitting the irrelative terms, the objective function in (\ref{DC1}) turns into:
\begin{equation}\label{sdr-main}
  \mathbf{v}^{H}\mathbf{\Phi}\mathbf{v}+\mathbf\Xi_{1}\mathbf{v}+\mathbf v^{H}\mathbf{\Xi_{2}},
\end{equation}
where $\mathbf\Xi_{1}=\mathbf D- \mathbf{D}^{'}$ and $\mathbf\Xi_{2}=\mathbf{D}^{H}-\mathbf{D}^{'H}$.

Consider that the problem (\ref{sdr-main}) is still a non-convex quadratically constrained quadratic program (QCQP), which can be reformulated as a homogeneous QCQP. By introducing an auxiliary variable t, satisfying $\|\text t\|^{2}=1$, (\textbf P1-1-3) can be expressed as:
\begin{align}
 \underset{\mathbf {\widetilde v}}{\mathrm {max}}  \quad &{\mathbf{\widetilde v}}^{H}{\mathbf{\Psi}}{\mathbf{\widetilde v}} \label{sdrmain1} \\
\mathrm{s.t.}  \quad   &||\mathbf {\widetilde v}_{n}||=1, n=1,2,\cdots,N+1. \tag{\ref{sdrmain1}{a}} \label{sdrmain1a}
\end{align}
where
\begin{align}\label{psi}
\mathbf{\Psi}=\left[\begin{array}{cc}
\mathbf{\Phi} & \mathbf{\Xi}_{2} \\
\mathbf{\Xi}_{1} & 0
\end{array}\right],
\end{align}
and
\begin{equation}\label{sdrsmp}
 \mathbf {\widetilde v}^{H} = [\mathbf v, t]^{H}.
\end{equation}

The objective function of (\ref{sdrmain1a}) can be rewritten as $\mathbf{\widetilde v}^{H}{\mathbf{\Psi}}\mathbf{\widetilde v}=\rm{tr} \left(\bm{{\mathbf{\Psi}}}\textbf{Q} \right)$ with $\textbf{Q}=\mathbf{{\widetilde v}{\widetilde v}}^H$.
In particular, $\textbf{Q}$ is a positive semi-infinite matrix with $\rm{rank}(\textbf{Q})=1$. Nevertheless, since the rank-one constraint is non-convex, we apply the semi-infinite relaxation (SDR) method to relax this constraint and reformulate (\textbf{P1-1-2}) as
\begin{align}\label{p1.2}
&\max ~\rm{tr} \left( \bm{{\mathbf{\Psi}}}\textbf{Q} \right)\\
&\nonumber \textrm{s.t.} \ ~~\textbf{Q}_{n,n}=1,n=1,2,\cdots,N+1\\
&\nonumber~~~~~~~\textbf{Q}\succeq0,
\end{align}
which is a convex semi-infinite programming (SDP) problem that can be solved by existing convex optimization solvers such as CVX \cite{MGCVX}. It is worth noting that the optimal solution $\textbf{Q}^*$ of the problem (\ref{p1.2}) may not be a rank-one solution after relaxation, which can be solved by Gaussian randomization \cite{8811733}.

\section{Proposed hybrid precoding methods}
In IRS-aided hybrid SM system, transmit precoding is an important technique to improve the system safety performance. In this section, two transmit precoding design methods are proposed for the IRS-aided secure hybrid SM system: ASR-SCA and COR-GA, respectively.

As the IRS beamformer vector $\bf{V}$ is fixed, \textbf{P1-2} can be rewritten as:
\begin{align}\label{}
\left( \textbf{P1-2} \right):	\underset{\mathbf p}{\mathrm {max}}  \quad &R^{a}_{s}  \label{trans2} \\
	\mathrm{s.t.}  \quad &||\mathbf P||\leq N_{RF} \tag{\ref{trans2}{a}}
\end{align}

Since the precoding matrix \textbf{P} is diagonal, it can be expressed in the vector form, i.e, \textbf{p}. Therefore, it is easy to convert the numerators in \textbf{P1-2} to the following forms:
\begin{align}\label{expand1}
&||\sqrt{\beta P_s}\mathbf\Omega_{B}^{-\frac{1}{2}}{\bf{(H+GVF)}}\mathbf D_{mn}\mathbf{P}||^2\\
\nonumber&=\beta P_s {\mathbf p^H}{\mathbf D_{mn}^H}{\mathbf {(H+GVF)}}^H{\mathbf\Omega_{B}^{-H}}{\mathbf {(H+GVF)}}{\mathbf D_{mn}}{\mathbf p}\\
\nonumber&=\beta P_s\cdot{\mathbf p^H}\mathbf B_{mn}{\mathbf p},
\end{align}
\begin{align}\label{expand2}
&||\sqrt{\beta P_s}\mathbf\Omega_{E}^{-\frac{1}{2}}{\bf{(Q+MVF)}}\mathbf D_{m^{'}n^{'}}\mathbf{P}||^2\\
\nonumber&=\beta P_s {\mathbf p^H}{\mathbf D_{m^{'}n^{'}}^H}{\mathbf {(Q+MVF)}}^H{\mathbf\Omega_{E}^{-H}}{\mathbf {(Q+MVF)}}{\mathbf D_{m^{'}n^{'}}}{\mathbf p}\\
\nonumber&=\beta P_s\cdot{\mathbf p^H}\mathbf E_{m^{'}n^{'}}{\mathbf p},
\end{align}
where
\begin{equation}
\mathbf B_{mn}={\mathbf D_{mn}^H}{\mathbf {(H+GVF)}}^H{\mathbf\Omega_{B}^{-H}}{\mathbf {(H+GVF)}}{\mathbf D_{mn}},
\end{equation}
\begin{equation}\label{}
\mathbf E_{m^{'}n^{'}}={\mathbf D_{m^{'}n^{'}}^H}{\mathbf {(Q+MVF)}}^H{\mathbf\Omega_{E}^{-H}}{\mathbf {(Q+MVF)}}{\mathbf D_{m^{'}n^{'}}}.
\end{equation}
As a consequence, the SR has a new expression as follows:
\begin{align}\label{fixv}
 R^{a}_{s}=&{\rm{log}_2}\sum_{m^{'}=1}^{N_{RF}M}\sum_{n^{'}=1}^{N_{RF}M}{\rm{exp}}\left(\frac{-\beta P_s\cdot{\mathbf p^H}\mathbf E_{m^{'}n^{'}}{\mathbf p}}{4}\right)-\\
 \nonumber &{\rm{log}_2}\sum_{m=1}^{N_{RF}M}\sum_{n=1}^{N_{RF}M}{\rm{exp}}\left(\frac{-\beta P_s\cdot{\mathbf p^H}\mathbf B_{mn}{\mathbf p}}{4}\right).
 \end{align}

\subsection{Proposed ASR-SCA method}
{
$\textbf{Algorithm 2}$
\begin{algorithm}
	\renewcommand{\algorithmicrequire}{\textbf{Input:}}
	\renewcommand{\algorithmicensure}{\textbf{Output:}}
	\caption{ProposedASR-SCA beamformer}
	\label{alg:1}
	\begin{algorithmic}[1]
		\REQUIRE the channel matrix $\textbf{H}$, $\textbf{Q}$, $\textbf{M}$, $\textbf{G}$, $\textbf{F}$, $\mathbf{V}$, and the $\mathcal{M}$-ary constellation
		\ENSURE $\mathbf {p}$
        \STATE Computing the upper bound of $R_{B}$ and the lower bound of $R_{E}$ as $R^{u}_{B}$ and $R^{l}_{E}$, respectively.
		\STATE Initialize $\mathbf {p}_0$ to a feasible value.
        \STATE Initialize step $k=0$
        \STATE Reformulate the problem SCA to get the convex objective function (\ref{trans1})
        \REPEAT
        \STATE Let $k=k+1$
        \STATE Update the procoding vector $\mathbf{p}$ by solving the reformulated convex optimization problem (\ref{trans1}) over $\mathbf{p}$  for fixed $\mathbf v$
        \UNTIL $\|\mathbf p_{k}-\mathbf p_{k-1}\|_2\leq0.01$
        \STATE \textbf{return} $\mathbf p$
	\end{algorithmic}
\end{algorithm}
It is observed in Eq.(\ref{trans2}) that it is a QCQP problem, which is difficult to solve directly. Therefore, in this section, this problem is transformed into a concave maximization problem through the method of SCA, and finally, an effective solution is obtained through optimization tools.

The corresponding average SR for Bob and Eve are denoted respectively as:
\begin{equation}\label{}
 R^{a}_{E}={\rm{log}_2}\sum_{m^{'}=1}^{N_{RF}M}\sum_{n^{'}=1}^{N_{RF}M}{\rm{exp}}\left(\frac{-\beta P_s\cdot{\mathbf p^H}\mathbf E_{m^{'}n^{'}}{\mathbf p}}{4}\right),
\end{equation}
\begin{equation}\label{}
 R^{a}_{B}={\rm{log}_2}\sum_{m=1}^{N_{RF}M}\sum_{n=1}^{N_{RF}M}{\rm{exp}}\left(\frac{-\beta P_s\cdot{\mathbf p^H}\mathbf B_{mn}{\mathbf p}}{4}\right).
\end{equation}

We take first-order Taylor expansion of $R^{a}_{E}$ at a given feasible point:
\begin{align}
{\rm{exp}}&\left(-\tau{\mathbf p^H}\mathbf E_{m^{'}n^{'}}{\mathbf p}\right)\\
>\nonumber&{\rm{exp}}\left(-\tau{\mathbf p^H_{0}}\mathbf E_{m^{'}n^{'}}{\mathbf p_{0}}\right)\cdot \\
\nonumber&[1+\left(\tau{\mathbf p^H_{0}}\mathbf E_{m^{'}n^{'}}{\mathbf p_{0}}\right)-\left(\tau{\mathbf p^H}\mathbf E_{m^{'}n^{'}}{\mathbf p}\right)],
\end{align}
where $\tau=\frac{\beta P_s}{4}$, and the lower bound $R^{l}_{E}$ of $R^{a}_{E}$ is expressed as:
\begin{align}\label{app-low}
  R^{a}_{E}>&R^{l}_{E}\\
  \nonumber&={\rm{log}_2}\sum_{m=1}^{N_{RF}M}\sum_{n=1}^{N_{RF}M}{\rm{exp}}\left(-\tau{\mathbf p^H_{0}}\mathbf E_{m^{'}n^{'}}{\mathbf p_{0}}\right)\cdot\\
  \nonumber&[1+\left(\tau{\mathbf p^H_{0}}\mathbf E_{m^{'}n^{'}}{\mathbf p_{0}}\right)-\left(\tau{\mathbf p^H}\mathbf E_{m^{'}n^{'}}{\mathbf p}\right)].
\end{align}

As $\mathbf E_{m^{'}n^{'}}$ is a semi positive definition matrix, it can be inferred that $R^{l}_{E}$ is a concave function.

Next, we seek the upper bound of $R^{a}_{B}$. As ${\mathbf p^H}\mathbf B_{mn}{\mathbf p}$ is a convex function with respect to $\mathbf p$, the first order Taylor expansion at a feasible point $\mathbf p_{0}$ is derived as:
\begin{equation}
{\mathbf p^H}\mathbf B_{mn}{\mathbf p}\geq 2\mathfrak{Re}\{{\mathbf p^H_{0}}\mathbf B_{mn}{\mathbf p}\}-{\mathbf p^H_{0}}\mathbf B_{mn}{\mathbf p_{0}}.
\end{equation}
Considering the coefficient $\tau$, we can get:
\begin{equation}
-\tau{\mathbf p^H}\mathbf B_{mn}{\mathbf p}\leq \tau{\mathbf p^H_{0}}\mathbf B_{mn}{\mathbf p_{0}}-2\tau\mathfrak{Re}\{{\mathbf p^H_{0}}\mathbf B_{mn}{\mathbf p}\}.
\end{equation}
And the upper bound $R^{u}_{B}$ of $R^{a}_{B}$ is exported as:
\begin{align}\label{app-up}
  R^{a}_{B}\leq&R^{u}_{B}\\
  \nonumber&={\rm{log}_2}\sum_{m=1}^{N_{RF}M}\sum_{n=1}^{N_{RF}M}{\rm{exp}}\big(\tau{\mathbf p^H_{0}}\mathbf B_{mn}{\mathbf p_{0}}\\
  \nonumber&~~-2\tau\mathfrak{Re}\{{\mathbf p^H_{0}}\mathbf B_{mn}{\mathbf p}\}\big).
\end{align}

Accordingly, the ASR for the hybrid SM system has the new expression:
\begin{equation}\label{}
  R^{n}_{s} = R^{l}_{E}-R^{u}_{B}
\end{equation}
\textbf{P1-2} can be rewritten as:
\begin{align}
	\underset{\mathbf p}{\mathrm {max}}  \quad &R^{n}_{s}(\mathbf p)  \label{trans1} \\
	\mathrm{s.t.}  \quad &||\mathbf P||\leq N_{RF} \tag{\ref{trans1}{a}}
\end{align}
}
It is easy to find that the objective function turns into the combination of a concave function and a linear function, and can be solved via existing convex optimization tools such as CVX.

\subsection{Proposed COR-GA method}
In the previous section, the ASR-SCA algorithm was presented to optimize the hybrid precoder for higher SR. For the comparison of  the secrecy performance and  to offer a new solution to
this non-convex optimization problem, we propose another secure transmit power scheme, namely COR-GA based on the cut-off rate-based SR expression, in what follows.
To maximize $R_s^a(\mathbf{p})$, the COR-GA method can be employed to directly optimize the hybrid precoding vector $\mathbf{p}$. We derive the gradient of $R_s^a(\mathbf{p})$  with respect to $\mathbf{p}$ as:
 \begin{align}\label{maxga}
&\frac{\partial R^{a}_{s}}{\partial \mathbf P}=\frac{\beta P_s}{4{\rm{ln}2}}\frac{\sum_{m=1}^{N_{RF}M}\sum_{n=1}^{N_{RF}M}\chi^{B}_{mn}(\mathbf{p})(\mathbf B_{mn}+\mathbf B_{mn}^{H}){\mathbf P}}{\sum_{m=1}^{N_{RF}M}\sum_{n=1}^{N_{RF}M}\chi^{B}_{mn}(\mathbf{p})}\\
\nonumber&-\frac{\beta P_s}{4{\rm{ln}2}}\frac{\sum_{m^{'}=1}^{N_{RF}M}\sum_{n^{'}=1}^{N_{RF}M}\chi^{E}_{m^{'}n^{'}(\mathbf{p})}(\mathbf E_{m^{'}n^{'}}+\mathbf E_{m^{'}n^{'}}^{H}){\mathbf p}}{\sum_{m^{'}=1}^{N_{RF}M}\sum_{n^{'}=1}^{N_{RF}M}\chi^{E}_{m^{'}n^{'}}(\mathbf{p})}
 \end{align}
where
\begin{equation}\label{chib}
  \chi^{B}_{mn}(\mathbf{p})={\rm{exp}}\left(\frac{-\beta P_s\cdot{\mathbf p^H}\mathbf B_{mn}{\mathbf p}}{4}\right),
\end{equation}
\begin{equation}\label{chie}
  \chi^{E}_{m^{'}n^{'}}(\mathbf{p})={\rm{exp}}\left(\frac{-\beta P_s\cdot{\mathbf p^H}\mathbf E_{m^{'}n^{'}}{\mathbf p}}{4}\right).
\end{equation}
In order to find a locally optimal  $\beta$, we first initialize $\mathbf{p}$ and $R_s^{a}(\mathbf{p})$,  solve the gradient $\nabla_{\mathbf p} R^{a}_{s}(\mathbf{p})$, and adjust $\mathbf{p}$ according to $\nabla_{\mathbf p} R^{a}_{s}(\mathbf{p})$. The value of $\mathbf{p}$ is updated according to the following iterative formula
\begin{equation}\label{update}
  \mathbf{p}_{k+1}=\mathbf{p}_{k}+\mu\nabla_{\mathbf p} R^{a}_{s}
\end{equation}
Then, obtain $R_s^{a}$ and update $\mathbf{p}$ or step size $\mu$ according to the difference between before and after $R_s^{a}$, and repeat the above
steps until the termination condition is reached.
\section{JOINT OPTIMIZATION SCHEME DESIGN}{
Based on the separate design methods of IRS beamforming and transmitter hybrid precoding described above, $\textbf{Algorithm 3}$ presents the overall alternate iterative algorithm for the objective function Eq. (\ref{main}). By continuously optimizing the IRS beamforming vector $\mathbf v$ and hybrid precoding vector $\mathbf p$, the SR value of the system will continue to increase until the convergence condition is reached.
\begin{algorithm}
	\renewcommand{\algorithmicrequire}{\textbf{Input:}}
	\renewcommand{\algorithmicensure}{\textbf{Output:}}
	\caption{Proposed Max-ASR-SCA beamformer}
	\label{alg:1}
	\begin{algorithmic}[1]
		\REQUIRE the channel matrix $\textbf{H}$, $\textbf{Q}$, $\textbf{M}$, $\textbf{G}$, $\textbf{F}$, $\mathbf{V}$, and the $\mathcal{M}$-ary constellation
        \STATE Initialize step $k=0$ and $\varepsilon$
        \REPEAT
        \STATE Obtain $\mathbf v_{k+1}$ via solving (\ref{fxp}) using IRS-ADMM or IRS-BCA or IRS-SDR with fixed $\mathbf p_k$
        \STATE Update $\mathbf v_{k+1} = \mathbf v_{k}$ if $R_s^l(\mathbf v_{k+1}, \mathbf p_{k})<R_s^l(\mathbf v_{k}, \mathbf p_{k})$
        \STATE Obtain $\mathbf p_{k+1}$ via solving (\ref{trans2}) using ASR-SCA or COR-GA with fixed $\mathbf v_{k+1}$
        \STATE Let $k=k+1$
        \UNTIL $\|R_s^l(\mathbf v_{k}, \mathbf p_{k})-R_s^l(\mathbf v_{k-1}, \mathbf p_{k-1})\|\leq\varepsilon$
        \STATE \textbf{return} $\mathbf v^* = \mathbf v$ and $\mathbf p^* = \mathbf p$
	\end{algorithmic}
\end{algorithm}
}
\section{Complexity Analysis}
{
The unit of computational complexity is floating-point operations (FLOPs), which is omitted for convenience in what follows.

For the proposed IRS-ADMM, the computational complexity is partitioned into three parts: (a)~the computation of the auxiliary variables $\mathbf{a}_{ij}$ and $\mathbf{s}_{ij}$ in Eq.(\ref{asij}), (b)~the computation of $\mathbf{B}^{mn}_{ij}$, $\mathbf{B}^{m'n'}_{i'j'}$, $\mathbf{C}^{mn}_{ij}$, $\mathbf{C}^{m'n'}_{i'j'}$, $\mathbf{D}$ and $\mathbf{D}^{'}$, (c)~the computation of the solution of the Lagrange function in Eqs.(\ref{admm1})-(\ref{admm3}).
For part (a), the computational complexity of $\mathbf{a}_{ij}$ and $\mathbf{s}_{ij}$ can be expressed as $8N_{b}N_{k}-2N_{b}$ and $8NN_{k}-2N$ respectively, and $\emph{C}_{a}$ denotes the complexity of this part.
The complexity of $\mathbf{B}^{mn}_{ij}$, $\mathbf{C}^{mn}_{ij}$ and $\mathbf{D}$ can be represented as $8N^{2}N_{b}-2N^{2}$, $8N^{2}N_{b}-2NN_{b}$ and $8NN_{b}-2N$ separately. For $\mathbf{B}^{m'n'}_{i'j'}$, $\mathbf{C}^{m'n'}_{i'j'}$ and $\mathbf{D}^{'}$, it can be inferred from system structure that they have a similar complexity as above by replacing $N_b$ with $N_e$, $\emph{C}_{b}$ denotes the complexity of this part.
And the solution of the Lagrange function consists of three steps, for (\ref{admm1}), considering the inverse operation and the complex multiplication, its computational complexity can be expressed as $\emph{C}_{\mathbf{u}}=N^{3}+8N^{2}-2N$. For (\ref{lamb1}), $\emph{C}_{\mathbf{\lambda}}=16N^{2}-4N$. For (\ref{admm3}), assuming that each element in the IRS beamforming vector requires constant mode operation, the maximum complexity is $N$.

As a result, the computational complexity of IRS-ADMM method is:
\begin{align}
  \emph{C}_{IRS-ADMM}=&(N_{RF}M)\emph{C}_{a}+(N_{RF}M)^{2}\emph{C}_{b}+\\
\nonumber&D_{ADMM}(N^{3}+24N^{2}-5N)
\end{align}
where $D_{ADMM}$ denotes the iteration time of the ADMM algorithm.

For the proposed IRS-BCA, the computational complexity is partitioned into three parts: (a)~the computation of the auxiliary variables $\mathbf{a}_{ij}$ and $\mathbf{s}_{ij}$ in Eq.(\ref{asij}), (b)~the computation of $\mathbf{B}^{mn}_{ij}$, $\mathbf{B}^{m'n'}_{i'j'}$, $\mathbf{C}^{mn}_{ij}$, $\mathbf{C}^{m'n'}_{i'j'}$, $\mathbf{D}$ and $\mathbf{D}^{'}$, (c)~the computation of the constant mode operation in Eq.(\ref{constant}). The calculation of (a)-(b) is omitted here because
it has been described. And the computational complexity required for (c) is $N$.

Hence, the computational complexity of the IRS-BCA method is:
\begin{align}
  \emph{C}_{IRS-BCA}=(N_{RF}M)\emph{C}_{a}+(N_{RF}M)^{2}\emph{C}_{b}+D_{BCA}N
\end{align}
where $D_{BCA}$ stands for the iteration time of the BCA algorithm.

For the proposed IRS-SDR, the computational complexity is partitioned into three parts: (a)~the computation of the auxiliary variables $\mathbf{a}_{ij}$ and $\mathbf{s}_{ij}$ in Eq.(\ref{asij}), (b)~the computation of $\mathbf{B}^{mn}_{ij}$, $\mathbf{B}^{m'n'}_{i'j'}$, $\mathbf{C}^{mn}_{ij}$, $\mathbf{C}^{m'n'}_{i'j'}$, $\mathbf{D}$ and $\mathbf{D}^{'}$, (c)~the computation of the problem in Eq.(\ref{p1.2}). The calculation of (a)-(b) is omitted here because
it has been described. And the complexity required for SDP problem Eq.(\ref{p1.2}) using CVX is $\emph{C}_{SDP}=\mathcal{O}\big(N^{4.5}{\rm log(\frac{1}{\varsigma})}\big)$.

Hence, the computational complexity of IRS-SDR method is:
\begin{align}
  \emph{C}_{IRS-SDR}=&(N_{RF}M)\emph{C}_{a}+(N_{RF}M)^{2}\emph{C}_{b}+\\
  \nonumber &D_{SDR}\mathcal{O}\big(N^{4.5}{\rm log(\frac{1}{\varsigma})}\big)
\end{align}
where $D_{SDR}$ represents the iteration time of the SDR algorithm.

It can be seen from the above analysis that the computational complexity of the IRS beamforming mainly depends on the number of reflection units. The complexity orders of IRS-BCA, IRS-ADMM and IRS-SDR are $\mathcal{O}(N^2)$, $\mathcal{O}(N^3)$ and $\mathcal{O}(N^{4.5})$ respectively.

For the proposed ASR-SCA method, ignoring the computational complexity of those fixed terms, the computational complexity is divided into three parts: (a)~the computation of (\ref{app-low}), (b)~the computation of (\ref{app-up}), (c)~the computation of the function (\ref{trans1}) using CVX.
For part (a), the computational complexity of $\bf{p}_{0}^{H}\bf{B}_{m^{'}n^{'}}\bf{p}_{0}$ and $\bf{p}^{H}\bf{B}_{m^{'}n^{'}}\bf{p}$ have the same value of $8(N_{RF}N_{K})^{2}+6N_{RF}N_{K}-2$, hence,
$R_{E}^{l}$ requires $\emph{C}_{R_{E}^{l}}=(N_{RF}M)^2(\emph{C}_{\bf{p}_{0}^{H}\bf{B}_{m^{'}n^{'}}\bf{p}_{0}}+\emph{C}_{\bf{p}^{H}\bf{B}_{m^{'}n^{'}}\bf{p}})= 2(N_{RF}M)^2(8(N_{RF}N_{K})^{2}+6N_{RF}N_{K}-2)$.

We can get the computational complexity of part (b) from (\ref{app-up}),
\begin{equation}
  \emph{C}_{R_{E}^{u}}=2(N_{RF}M)^2(8(N_{RF}N_{K})^{2}+6N_{RF}N_{K}-2).
\end{equation}
}
The highest-order computational complexity per iteration in part (b) is $\mathcal{O}((N_{RF}N_{K})^3)$.
Therefore, the computational complexity of ASR-SCA method is:
\begin{align}
  \emph{C}_{SCA}=&D_{SCA}\Big(4(N_{RF}M)^2\\
\nonumber    &\big(8(N_{RF}N_{K})^{2}+6N_{RF}N_{K}-2)+(N_{RF}N_{K})^3\big)\Big),
\end{align}
  where $D_{SCA}$ denotes the iteration time of the SCA algorithm.

For the proposed COR-GA method, the computational complexity is composed of three parts: (a)~the computation of $\chi_{mn}^{B}$ and $\chi_{m^{'}n^{'}}$ in Eqs. (\ref{chib}) and (\ref{chie}), (b) the computation of the second part of the numerator of Eq. (\ref{maxga}), (c) the computation of    Eq.(\ref{update}).
The computational complexity of $\chi_{mn}^{B}$ requires $\emph{C}_{\chi_{mn}^{B}}=8(N_{RF}N_{K})^{2}+6N_{RF}N_{K}-2$, and the computation of $\chi_{mn}^{B}$ and $\chi_{m^{'}n^{'}}$ are the same, therefore, the complexity of part (a) is expressed:
\begin{equation}
  \emph{C}_{a}=16(N_{RF}N_{K})^{2}+8N_{RF}N_{K}-4
\end{equation}
For part (b), the computation complexity of the second part of the numerator of Eq. (\ref{maxga}) is expressed as \\
$\emph{C}_{b}=16(N_{RF}N_{k})^2-4N_{RF}N_{k}$.
And the computational complexity of the update function in Eq.(\ref{update}) requires $\emph{C}_{c}=6N_{RF}N_{k}$.
Consequently, the computational complexity of COR-GA method is:
\begin{align}
  \emph{C}_{GA}=&D_{GA}\Big((N_{RF}M)^{2}\\
\nonumber  &\big(32(N_{RF}N_{K})^{2}+4N_{RF}N_{K}-4\big)+6N_{RF}N_{k}\Big)
\end{align}
where $D_{GA}$ stands for the iteration time of the GA method.

Concerning the hybrid precoding algorithms, as the number of RF chains $N_{RF}$ and the number of transmit antenna in each sub-array are the main impact factor, we can see that the proposed ASR-SCA method has higher computational complexity than COR-GA.

The complexity of the joint optimization schemes comes from IRS beamforming and transmit hybrid precoding. Assuming that the number of iterations of the outer layer  is $D_{OUT}$, the computational complexity of the joint optimization scheme is
\begin{equation}\label{}
  \emph{C}_{OUT}=D_{OUT}(\emph{C}_{IRS}+\emph{C}_{PRE})
\end{equation}
where $\emph{C}_{IRS}$ and $\emph{C}_{PRE}$ denote the complexity of the selected IRS beamforming algorithm and the complexity of the selected precoding algorithm, respectively.

\section{Simulation Results}
\begin{figure}[ht]
\centerline{\includegraphics[width=0.48\textwidth]{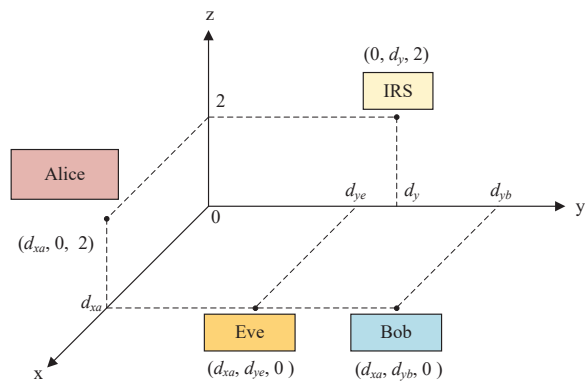}}
\caption{3D coordinate diagram of the IRS-aided hybrid SSM system.}
\label{coordinate}
\end{figure}

In this section, a three-dimension coordinate diagram shown in Fig.~\ref{coordinate} is considered, where the coordinates of Alice, IRS, Bob, and Eve are ($d_{xa}$m, 0m, 2m), (0m, $d_y$m, 2m), ($d_{xb}$m, $d_{yb}$m, 0m), ($d_{xe}$m,  and $d_{ye}$m, 0m), respectively. Assume that all channels follow the Rayleigh fading model \cite{5281762,6094142} and the path loss at distance \emph{d} is:
\begin{equation}
  PL(d)=PL_0-10\alpha\text{log}_{10}(\frac{d}{d_0}),
\end{equation}
where $PL_0=-30\text{dB}$ represents the path loss at a distance of $d_0=1\text{m}$ and $\alpha$ denotes path loss exponent.
The path loss exponents of the Alice-IRS, Alice-Bob/Eve and IRS-Bob/Eve are set to be 2.2, 2.7 and 2.5 respectively.
In IRS-assisted communication scenarios, placing IRS at the transmitter or receiver side helps to improve the security performance of the system. Therefore, we place IRS near the receiver and the parameters are set as $d_{xa}=d_{xb}=d_{xe}=10, d_{yb}=45$ and $d_{ye}=35$. And quadrature phase shift keying (QPSK) modulation is employed. The other simulation parameters are set as: $N_{RF}=8$, $N_k=4$,
$\beta = 0.35$, $P_s=30$dBm and $N = 50$. For the sake of fairness of Bob and Eve, it is assumed that they have the same receive ability and all noise variances in channels are identical, i.e., $N_b=N_e=2$ and $\sigma_b^2=\sigma_e^2=-80$dBm.

\begin{figure}
  \centering
  \includegraphics[width=0.45\textwidth]{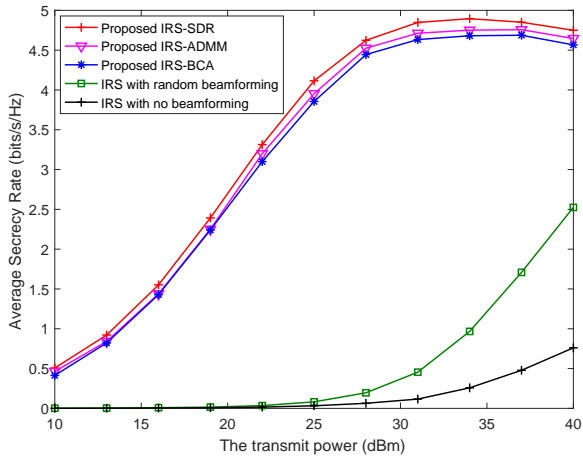}\\
  \caption{Curves of average SR versus transmit power for different IRS beamforming algorithms.}
\label{IRS-OPT}
\end{figure}

The average SR for the different IRS beamforming algorithms is illustrated in Fig.~\ref{IRS-OPT} when the hybrid precoding is fixed.
It can be seen from the figure that the safety performance of all three proposed algorithms is better than that of the random phase scheme, which illustrates the importance of the IRS beamforming design. Furthermore, the safety performance of IRS-SDR is significantly better than those of IRS-ADMM and IRS-BCA. As the transmit power increases, IRS-SDR approaches the upper bound of the achievable safety rate faster than the other two algorithms. The proposed IRS-SDR algorithm outperforms the IRS-ADMM and IRS-BCA algorithms in terms of safety performance.

\begin{figure}
  \centering
  \includegraphics[width=0.45\textwidth]{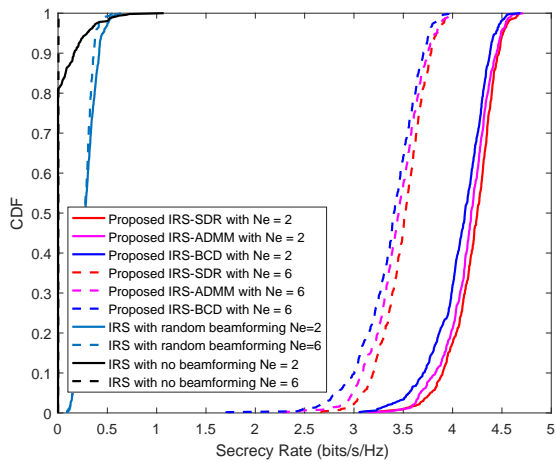}\\
  \caption{Curves of CDF versus SR achieved by different IRS beamforming algorithms for different values of Ne with $P_t$=30dBm.}
\label{IRS-CDF}
\end{figure}

Fig.~\ref{IRS-CDF} shows the cumulative distribution function (CDF) of the three IRS beamforming algorithms for different numbers of Eve's antennas when $P_t = 30$dBm. It is seen from the figure that the CDF curves for all proposed beamforming algorithms shift to the left as the number of Eve's antennas increases, indicating that the SR performance decreases as the eavesdropping capability improves. From Fig.~\ref{IRS-CDF}, the same descending ranking of the proposed IRS beamforming algorithms is seen as Fig.~\ref{IRS-OPT}: IRS-SDR, IRS-ADMM, and IRS-BCA.

\begin{figure}
  \centering
  \includegraphics[width=0.45\textwidth]{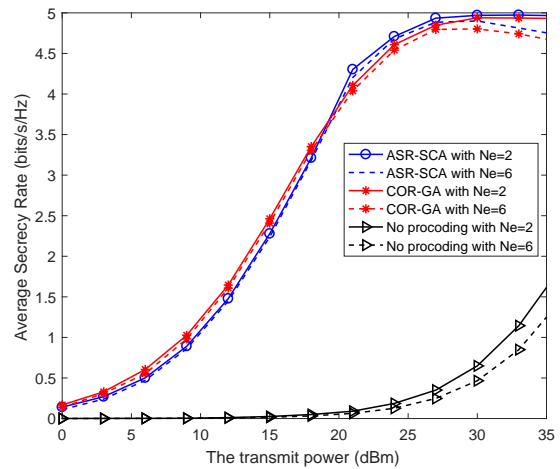}\\
  \caption{Curves of average SR versus transmit power of different hybrid precoding algorithms for different values of Ne.}
\label{precoding}
\end{figure}

In order to investigate the SR performance of the proposed two hybrid precoding algorithms proposed, the SR for different numbers of EVE's antennas with IRS applied IRS-BCA beamforming algorithm are presented in Fig.~\ref{precoding}. From this figure, it can be seen that the SR performance of the system is greatly improved with precoding compared to the case without the design of hybrid precoding. The upper bound of SR achieved by both algorithms decrease by 0.1bits/s/Hz when the number of EVE's antennas increased from 2 to 6. This is due to the fact that with a high SNR, the channel condition is good at both the desired user end and the eavesdropping user end, while the number of eavesdropping antennas is higher than that of the desired side, resulting in SR performance decrease. With the same IRS beamforming, the SR performance of ASR-SCA is lower than that of COR-GA at lower transmit power and better than that of COR-GA at higher transmit power. The reason is that at low transmit power, the eavesdropper is set closer to the transmitter, which results in a better channel condition than the legal end, and ASR-SCA utilizes the difference between the upper and lower bounds of the mutual information of the legitimate and eavesdropping users as the objective function, which causes the amount of mutual information at the eavesdropping end to be greater than that at the legitimate end, and the SR performance is relatively low. In the case of high transmit power, the channel condition of the legitimate user is also good. With IRS assisted, ASR-SCA can obtain an approximate optimal solution, while the gradient ascent based COR-GA usually converges to a local optimal solution, therefore its SR performance is not as good as ASR-SCA.

\begin{figure}
  \centering
  \includegraphics[width=0.45\textwidth]{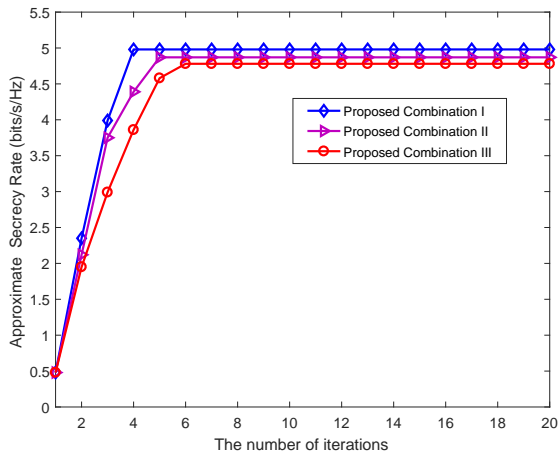}\\
  \caption{Curves of approximate SR versus number of iterations for different joint optimization algorithms.}
\label{cov}
\end{figure}

Fig.~\ref{cov} plots the approximate SR of the three jointly optimized IRS beamformer and transmit hybrid precoding schemes changes as the number of iteration increases. To balance the computational complexity of joint optimization schemes, there are three combinations for beamformer and precoding: 1)Combination I: IRS-BCA plus ASR-SCA; 2)Combination II: IRS-SDR plus COR-GA; 3)Combination III: IRS-ADMM plus COR-GA. In order to compare the performance more clearly, the same optimization starting point is selected for the three combinations. From Fig.~\ref{cov}, it is seen that at the start of the iteration, the approximate SR values of the three joint optimization schemes change significantly as the iterations continue, and as the number of iterations increases, the growth rate of approximate SR of the proposed schemes tends to flat until the convergence is reached. It can also be seen from the figure that all proposed schemes converge within 10 iterations, which demonstrates the advantages of the proposed schemes. In addition, comparing these three schemes, it can be seen that the convergence rate of Combination I is faster than those of Combinations II and III. Considering the complexity analysis in the previous section, we can see that ASR-SCA has a higher computational complexity than that of COR-GA, which means that Combination I trades a higher computational complexity for a less number of iterations, while the other two schemes sacrifice a certain number of iterations for a reduction in complexity.

\begin{figure}
  \centering
  \includegraphics[width=0.45\textwidth]{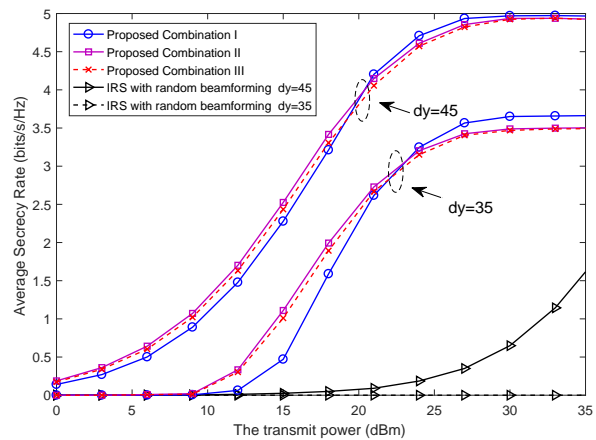}\\
  \caption{Curves of average SR versus transmit power of different joint optimization algorithms for different positions of IRS.}
\label{var-dis}
\end{figure}

Fig.~\ref{var-dis} shows the average SR versus transmit power for the joint optimization schemes when the IRS is placed at $d_y=45$ and $d_y=36$, i.e. close to the legal user and the eavesdropping user, respectively. The figure compares the proposed scheme with one where the IRS employs random phase and no hybrid precoding at the transmitter. From Fig.~\ref{var-dis}, it can be observed that the SR performance of Combination I is more sensitive to poor channel conditions, such as low transmit power or IRS deployed closer to the eavesdropper, leading to a worse SR performance than the other two combinations. However, when the channel condition is favorable, Combination I demonstrates a superior SR performance, highlighting the high channel demands of the ASR-SCA precoding design method. Additionally, when the IRS is placed near the eavesdropper, the SR performance suffers a lot. This implies that the proper IRS position configuration is essential to maintain optimum safety performance.
\begin{figure}
  \centering
  \includegraphics[width=0.45\textwidth]{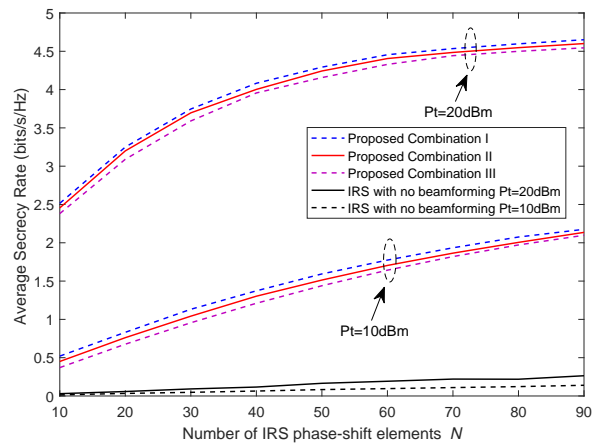}\\
  \caption{Curves of average SR versus number of IRS phase-shift elements for different joint optimization algorithms with different transmit power.}
\label{IRS-items}
\end{figure}

As depicted in Fig.~\ref{IRS-items}, the impact of the number of IRS phase-shift elements on joint optimization schemes with different transmit power is shown. It can be seen from the figure that the SR performances of the three proposed combinations improve as the transmit power increases. Furthermore, the SR performance of the proposed combinations is significantly higher that that of the case without beamforming, at both $P_t=10$dBm and $P_t=20$dBm conditions. And with the increase of the number of phase-shift elements, the SR of all combinations is increasing and is much better than that of no beamforming case. It is also worth noting that the proposed optimization solution exhibits even higher SR than that of the IRS without beamforming, despite the fact that the number of phase-shift elements is only 20.

\section{Conclusion}
In this paper, we have made investigated of IRS beamforming and transmit hybrid precoding design concerning IRS-aided hybrid SSM. In such an architecture, the first part of the bitstream is transmitted by APM symbol, and the second part of the bitstream is carried by selecting a transmit antenna subarray in the partially-connected structure rather than a single transmit antenna. Considering the physical-layer security, a cut-off rate based approximate SR expression was proposed.
Three beamforming design algorithms, IRS-ADMM, IRS-BCA, and IRS-SDR, have been proposed for IRS beamforming design. Simulation results showed that the proposed beamforming methods have an ascending order in SR: IRS-BCA, IRS-ADMM, IRS-SDR. For the transmit hybrid precoder, two hybrid precoding methods were also proposed: ASR-SCA and COR-GA. Simulation results showed that the proposed precoding methods have an ascending order in SR in the high transmit power region: no precoding scheme, COR-GA, ASR-SCA. Accordingly, three joint optimization schemes were given in this paper. The simulation results showed that compared with the schemes with no IRS beam optimization and no transmit precoding, the three proposed schemes can effectively improve the security performance of the system.

\bibliography{refr}
\bibliographystyle{IEEEtran}
\end{document}